\documentclass{jfm}

\usepackage{graphicx}
\usepackage{natbib}
\usepackage{latexsym}

\ifCUPmtlplainloaded \else
  \checkfont{eurm10}
  \iffontfound
    \IfFileExists{upmath.sty}
      {\typeout{^^JFound AMS Euler Roman fonts on the system,
                   using the 'upmath' package.^^J}%
       \usepackage{upmath}}
      {\typeout{^^JFound AMS Euler Roman fonts on the system, but you
                   dont seem to have the}%
       \typeout{'upmath' package installed. JFM.cls can take advantage
                 of these fonts,^^Jif you use 'upmath' package.^^J}%
       \providecommand\upi{\pi}%
      }
  \else
    \providecommand\upi{\pi}%
  \fi
\fi

\ifCUPmtlplainloaded \else
  \checkfont{msam10}
  \iffontfound
    \IfFileExists{amssymb.sty}
      {\typeout{^^JFound AMS Symbol fonts on the system, using the
                'amssymb' package.^^J}%
       \usepackage{amssymb}%

      }{}
  \fi
\fi

\ifCUPmtlplainloaded \else
  \IfFileExists{amsbsy.sty}
    {\typeout{^^JFound the 'amsbsy' package on the system, using it.^^J}%
     \usepackage{amsbsy}}
    {\providecommand\boldsymbol[1]{\mbox{\boldmath $##1$}}}
\fi

\def\cm{\nobreak\mbox{$\;$cm}}
\def\mm{\nobreak\mbox{$\;$mm}}
\def\m{\nobreak\mbox{$\;$m}}
\def\mPas{\nobreak\mbox{$\;$mPa.s}}
\def\mNm{\nobreak\mbox{$\;$mN\,m$^{-1}$}}
\def\kgm{\nobreak\mbox{$\;$kg\,m$^{-3}$}}
\def\Pa{\nobreak\mbox{$\;$Pa}}
\def\ms{\nobreak\mbox{$\;$m\,s$^{-1}$}}
\def\cms{\nobreak\mbox{$\;$cm\,s$^{-1}$}}

\providecommand\bnabla{\boldsymbol{\nabla}}

\newcommand\Rey{\mbox{\textit{Re}}}  
\newcommand\Web{\mbox{\textit{We}}}   
\newcommand\Webm{\mbox{\textit {We$^{\ast}$}}} 
\newcommand\Webmc{\mbox{\textit {We$^{\ast}_c$}}} 
\newcommand\Reym{\mbox{\textit {Re$^{\ast}$}}} 
\newcommand\Capi{\mbox{\textit {Ca}}}   

\newcommand\onB{\mbox{1/{\textit B}}} 
\newcommand\onBc{\mbox{1/{\textit{ B$_c$}}}}
\newcommand\onBm{\mbox{1/{\textit B'}}} 

\newsavebox{\astrutbox}
\sbox{\astrutbox}{\rule[-5pt]{0pt}{20pt}}

\title[Inertial effects in the Saffman-Taylor instability]{Inertial effects in the Saffman-Taylor instability}

\author[C. Chevalier, M. Ben Amar, D. Bonn and A. Lindner]%
{C\ls H\ls R\ls I\ls S\ls T\ls O\ls P\ls H\ls E\ns C\ls H\ls E\ls V\ls A\ls L\ls I\ls E\ls R$^{1,3}$,\break M\ls A\ls R\ls T\ls I\ls N\ls E\ns B\ls E\ls N\ns A\ls M\ls A\ls R$^2$,\ns D\ls A\ls N\ls I\ls E\ls L\ns B\ls O\ls N\ls N$^{2,4}$\break \and A\ls N\ls K\ls E\ns L\ls I\ls N\ls D\ls N\ls E\ls R$^1$
}

\affiliation{$^1$Laboratoire de Physique et M\'ecanique des Milieux H\'et\'erog\`enes,\\ UMR 7636 CNRS, Ecole Sup\'erieure de Physique et de Chimie Industrielles,\\ 10 rue Vauquelin, 75231 Paris cedex 05, France\\[\affilskip]
$^2$Laboratoire de Physique Statistique, UMR 8550 CNRS, Ecole Normale Sup\'erieure,\\ 24 rue Lhomond, 75231 Paris cedex 05, France\\[\affilskip]
$^3$Ecole Nationale des Ponts et Chauss\'ees, 6-8 avenue Blaise Pascal, Cit\'e Descartes,\\
Champs sur Marne, 77455 Marne-la-Vall\'ee cedex 2, France\\[\affilskip]
$^4$Van der Waals-Zeeman Institute, University of Amsterdam,\\ Valckenierstraat 65, 1018 XE Amsterdam, the Netherlands }

\pubyear{2005}
\pagerange{1--13}
\begin{document}

\maketitle

\begin{abstract}
For the Saffman-Taylor instability, the inertia of the fluid may become important for large Reynolds numbers \Rey. We investigate the effects of inertia on the width of the viscous fingers experimentally. We find that, due to inertia, the finger width can increase with increasing speed, contrary to what happens at small \Rey. We find that inertial effects need to be considered above a critical Weber number \Web. In this case it can be shown that the finger width is governed by a balance between viscous forces and inertia. This allows us to define a modified control parameter \onBm, which takes the corrections due to inertia into account; rescaling the experimental data with \onBm, they all collapse onto the universal curve for the classical Saffman-Taylor instability. Subsequently, we try and rationalize our observations. Numerical simulations taking into account a modification of Darcy law to include inertia, are found to only qualitatively reproduce the experimental findings, pointing to the importance of three-dimensional effects.
\end{abstract}

\section{Introduction}

Viscous fingering has received much attention as an archetype of pattern-formation problems and as a limiting factor in the recovery of crude oil \cite*[see][]{Saff58, Bens86, Homs87, Coud91}. Viscous fingers form when in a thin linear channel or Hele-Shaw cell, a fluid pushes a more viscous fluid. The interface between the fluids develops an instability leading to the formation of finger-like patterns. The viscous fingering instability has been studied intensively over the past few decades both theoretically and experimentally.

For the classical Saffman-Taylor instability the width of the finger is governed by the competition between viscous and capillary forces: viscous forces tend to narrow the finger whereas capillary forces tend to widen it. When air pushes a viscous fluid, as is usually the case, the relative finger width is thus determined by the capillary number $\Capi=\eta U / \gamma$, (with $\eta$ the viscosity, $U$ the velocity and $\gamma$ the surface tension) the ratio between viscous and capillary forces. In the vast majority of cases that have been studied so far, inertial forces are negligible. The importance of inertia is given by the relative importance of the inertial and viscous forces, quantified by the Reynolds number $\Rey=\rho Ub/\eta$, with $\rho$ the fluid density, and $b$ the plate spacing of the Hele-Shaw cell in which the experiments are conducted. In most studies of the instability, $b$ is small, and the fluids considered both in applications as well as in experimental studies are typically high-viscosity oils. This automatically leads to small Reynolds numbers ($\Rey<<1$), so that inertial effects may be neglected.

More recently viscous fingering has been studied in non-Newtonian fluids using for example polymer solutions \cite*[see][]{Smit92, Bonn95, Lind00, Vlad00, Kawa01,  Lind02}. For the dilute polymer solutions used in a number of these studies, the shear viscosity of the water-based solutions is typically close to the water viscosity, and consequently the Reynolds number may -and does- become larger than unity. This means that inertia may become important, and needs to be disentangled from the observed non-Newtonian flow effects. Also recently, corrections to the so called Darcy's law have been developed incorporating inertial effects \cite*[see][]{Gond97, Ruye01}.  Darcy's law relates the pressure gradient to the fluid velocity and is one of the fundamental equations of the Saffman-Taylor instability; if inertial corrections can simply be included in a modified Darcy's law, this would greatly facilitate the understanding of the effect of inertia on the instability. These recent developments suggest that a better understanding of the Newtonian fingering instability for high Reynolds numbers is both necessary and feasible. 

In this paper we explore the Saffman-Taylor instability for Newtonian fluids for Reynolds numbers up to $\Rey=100$. To do so, we use low-viscosity silicone oils, pushed by air. The paper is organized as follows. In Sec. 2 we will recall the basic equations for the Saffman-Taylor instability and introduce the corrections due to inertia. Sec. 3 describes the set-up and experimental methods. In Sec. 4 the experimental results concerning the finger widths as well as the validity of Darcy's law are presented and discussed. In Sec.~5 we will introduce some theoretical elements as well as numerical simulation and their comparison to the experimental results. Sec. 6 gives a summary of the obtained results.

\section{Theory and equations}

\subsection{Presentation and review of classical Saffman-Taylor instability}

\begin{figure}
  \begin{center}
  \includegraphics[height=3.5cm]{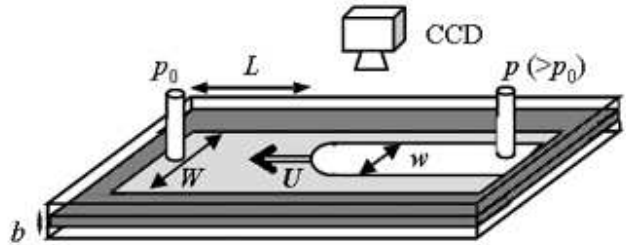}
  \caption{Schematic drawing of the experimental set-up.}
  \label{HS}
  \end{center}
\end{figure}

We study the Saffman-Taylor instability in a thin linear channel or Hele-Shaw cell (see figure \ref{HS}). The width of the cell $W$ is chosen to be large compared to the channel thickness $b$ and we thus work with high aspect ratios $W/b$. The cell is filled with a viscous fluid which is subsequently pushed by air. The viscosity and the density of air will be neglected throughout the paper.

When air pushes the viscous fluid due to an imposed pressure gradient $\nabla P$, an initially flat interface between the two fluids destabilizes. This destabilisation leads to the formation of a so called viscous finger; in steady state a stationary finger of width $w$ propagating at a velocity $U$ is found to occupy a fraction of the cell width: the relative finger width is defined as: $\lambda =w/W$.

For Newtonian fluids, the motion of a fluid in the Hele-Shaw cell is described by the two-dimensional velocity field $\boldsymbol{u}$ averaged through the thickness of the cell. It is given by Darcy's law, which relates the local pressure gradient to the velocity within the fluid as:
\begin{equation}
  \boldsymbol{u}=-\frac{b^2}{12\eta}\bnabla p.
  \label{Darcy1}
\end{equation}

It follows immediately that, if the fluid is incompressible, the pressure field satisfies Laplace's equation:
\begin{equation}
  \Delta p = 0.
  \label{LaplP}
\end{equation}

The pressure field is calculated within the driven fluid with in addition a pressure jump over the interface due to the surface tension: 
\begin{equation}
  \delta p=\gamma / R,
  \label{DeltP}
\end{equation}
with $R$ the radius of curvature of the interface, again employing a two-dimensional approximation, as was justified in the limit of small capillary numbers by \cite{Park84}, and \cite{Rein85}.

The other boundary conditions are the continuity condition, which implies that the normal velocity at both sides of the interface is equal and a far-field value for the pressure. Supplemented with these boundary conditions, (\ref{Darcy1}), (\ref{LaplP}) and (\ref{DeltP}) constitute the complete set of equations one can solve in order to obtain the complete finger shape for a given pressure gradient; and thus also its width.

For characterizing the instability quantitatively most studies have focused on the width of the finger $w$ relative to the channel width $W$: $\lambda =w/W$ as a function of their velocity. It follows from the above that their width is determined by the capillary number; one thus anticipates that the relative width of the viscous fingers decreases with increasing finger velocity.  This is indeed what is generally observed experimentally; in addition, for very large values of \Capi, $\lambda$ does not go to zero but reaches a limiting value of about half the channel width.
It also follows from the boundary conditions and (\ref{Darcy1}) to (\ref{DeltP}), that the control parameter for the fingering problem is $\onB = 12(W/b)^2 \Capi$ with $W/b$ the aspect ratio of the Hele-Shaw cell. When scaled on \onB, measurements of $\lambda$ for different systems all fall on the same universal curve. In the ideal, Newtonian, two-dimensional situation \onB \ is consequently the only parameter that determines the finger width \cite[see][]{Saff58, McLe81, Comb86, Hong86, Shra86}.

\subsection{Corrections due to inertia}
\label{secInertia}

When inertial forces have to be taken into account, both the Reynolds number $\Rey = \rho Ub/\eta$  or the Weber number $\Web = \rho U^2b/\gamma$ (the ratio of inertial forces to capillarity forces) may become important. We will now discuss how corrections due to inertia can be included in the basic equations.
 
Modifications of Darcy's law have first been proposed by \cite{Gond97} for parallel flow in a Hele-Shaw cell: they establish corrections by averaging inertia in the third dimension, i.e., they average over the direction of the plate spacing $b$, allowing them to derive a new nonlinear two-dimensional equation for the velocity field. \cite{Ruye01} suggests an improved correction starting from the three-dimensional Navier-Stokes equation. Inertial corrections are introduced in a perturbative fashion; using in addition a polynomial approximation to the velocity field, Ruyer-Quil proposes the modified two-dimensional Darcy's law of the form:
\begin{equation}
  \rho \left( \alpha \frac{\partial\boldsymbol{u}}{\partial t}+\beta\boldsymbol{u}\bnabla \boldsymbol{u}\right)=-\bnabla p-\frac{12\eta}{b^2}\boldsymbol{u},
  \label{Darcy2}
\end{equation}
with $\alpha=6/5$ and $\beta=54/35$. \cite{Plou02} also calculated inertial corrections to Darcy's law and arrive at a similar type of equation, but with slightly different coefficients. This equation leads to a better agreement between the linear stability analysis and the experimental data of Gondret \& Rabaud for the Kelvin-Helmholtz instability up to not too large Reynolds numbers. $\alpha$ and $\beta$ may vary depending on the way the averaging in the third dimension is done, but are always of order of 1.

When scaling length on $W$, time on $W/U$ and pressure on $12\eta UW/b^2$ one finds the following dimensionless equation:
\begin{equation}
  \Reym \left( \alpha\frac{\partial\boldsymbol{u}^\ast}{\partial t^\ast}+\beta\boldsymbol{u}^\ast\bnabla ^\ast\boldsymbol{u}^\ast\right)=-\bnabla^\ast p^\ast-\boldsymbol{u}^\ast,
  \quad \mbox{where\ }\quad
  \Reym=\frac{1}{12}\frac{b}{W}\frac{\rho U b}{\eta}=\frac{b}{12W}\Rey.  
  \label{Darcy2adim}
\end{equation}
\Reym \ is a modified Reynolds number, in the same way as the classical control parameter of the Saffman-Taylor instability \onB \ is a modified capillary number.

We can also introduce another number describing the importance between inertia and capillarity in the geometry of the Hele-Shaw, a modified Weber number:
\begin{equation}
\Webm=\frac{\rho U^2 W}{\gamma}=\frac{W}{b}\Web. 
  \label{Webmod}
\end{equation}

One important remark is that if one considers stationary and spatially uniform flow in our Hele-Shaw cell, it follows from (\ref{Darcy2}) that there are no corrections due to inertia, since $\partial \boldsymbol{u}/\partial t$ and $\boldsymbol{u}\bnabla \boldsymbol{u}$ are both zero. This will be the case in our fingering experiments far away from the moving interface and leads to the classical Darcy's law; we thus anticipate that it might remain valid even for relatively high \Rey.

\section{Experimental}

We use a linear Hele-Shaw cell consisting of two glass plates separated by a thin Mylar spacer. The plates are horizontal and clamped together in order to obtain a regular thickness $b$ of the channel. The thickness of the glass plates is chosen to be 2\cm \ in order to avoid any bending of the plates. The aspect ratio of the channel can be varied; we worked with different plate spacings $b$ and widths of the cell $W$, the length of the channel being always 1\m. The cell is filled with silicone oil and compressed air is used as the less viscous driving fluid.  

The silicone oils used were Rhodorsil 47V05, 47V10, 47V20 and 47V100 from Rhodia Silicones. Rheological measurements on a Reologica Stress-Tech rheometer confirmed the values of the viscosities $\eta$ of 5, 10, 20 and 100\mPas \ respectively, with no deviations larger than 4\%. We also used 47V02, its viscosity was measured to be 2.8\mPas. The surface tension $\gamma$ and the density $\rho$ of the silicone oils are 19.5~$\pm1$\mNm \ and 0.95~$\pm0.03$~$10^{-3}$\kgm \ as given by Rhodia Silicones.

The fingers were driven by applying a constant pressure drop $\Delta p=p_i-p_o$ between the inlet and the outlet of the cell. Depending on the order of magnitude of the applied pressure drop two methods were used. For $\Delta p$ larger than 3000\Pa \ we used compressed air and a pressure transducer at the entrance of the cell to fix $p_i$ at the inlet of the cell. In this case, the outlet was maintained at atmospheric pressure $p_o = p_{atm}$ due to an oil reservoir coupled to the cell. For a $\Delta p$ smaller than 3000\Pa, we obtained the pressure drop by lowering the oil reservoir at the outlet of the cell by a given amount, determining in this way $p_o$. In this case the inlet was maintained at atmospheric pressure $p_i = p_{atm}$.

The fingers were captured by a CCD camera, coupled to a data acquisition card (National Instruments) and a computer. This allowed for measurements of the relative width $\lambda  = w/W$ as a function of the velocity $U$ of the finger tip. For each configuration (cell geometry and fluid viscosity) several experimental runs (between 10 and 20) were performed increasing the applied pressure drop and thus the finger velocity until destabilization of the finger occurred; all the finger widths reported here correspond to stable fingers. 

In order to access high Reynolds numbers we can not only vary the velocity of the finger and the viscosity of the fluid but also change the thickness of the channel. We have thus worked with different channel geometries that are summarized on table \ref{dimen}.
\begin{table} 
  \begin{center} 
  \begin{tabular}{lcc} 
   & Thickness $b$ &	Width $W$ \\
  Geometry 1	& 0.25\mm	& 40\mm \\
  Geometry 2	& 0.75\mm	& 80\mm \\
  Geometry 3	& 0.75\mm	& 40\mm \\
  Geometry 4	& 1.43\mm	& 40\mm \\
\end{tabular}
  \caption{Different cell geometries used in our experiments.}
  \label{dimen}
  \end{center} 
\end{table} 
 
For the geometries used, the aspect ratio $W/b$ varies from 28 (geometry 4) to 160 (geometry 1). Even if an aspect ratio of 28 is rather small it is sufficient to consider the experiment as being quasi two-dimensional. It is observed that the results obtained for the high viscosity fluids (and thus a situation where inertial effects can be neglected) show only very little difference in their finger widths. The small difference is due to film effects \cite[see][]{Tabe86}, as will be discussed in more detail below.

Experiments were performed in all geometries for the silicon oils 47V05, 47V10 and 47V20. The Silicon oil 47V02 was tested in geometries 2 and 3, whereas the silicon oil 47V100 was used in geometries 3 and 4.
Finally, note that typical values of capillary number \Capi \ in the experiments are between 0.01 (for V02) and 0.5 (for V100).

\section {Presentation of the results}

\subsection{Darcy's law}

Assuming the flow far away from the finger to be uniform, we expect that the classical Darcy's law linking the gap averaged fluid velocity $V$ to the imposed pressure gradient $\nabla P$ in our Hele-Shaw cell remains valid for all of our experiments:
\begin{equation}
  V=-\frac{b}{12\eta}\nabla P.
  \label{DarcyMoy}
\end{equation}

Mass conservation allows to obtain the velocity $V$ of the fluid far away from the interface from the measured finger velocity $U$ simply by using $V=\lambda  U$, if one neglects the thin wetting film left on the glass plates behind the finger. The imposed pressure gradient is calculated by: $\nabla P = \Delta p / L$ where $\Delta p$ is the measured applied pressure drop and $L$ the distance between the finger tip and the exit of the cell.
 
In our experiments we reach high finger velocities and thus high capillary numbers \Capi. The influence of the thin wetting film left on the plates may therefore become important and can not be neglected any more in some of the experiments. It is taken into account using $V=\lambda U (1-2t/b)$, where $t$ is the thickness of the wetting film, which we estimate using the empirical result of \cite{Tabe86} and \cite{Tabe87}:
\begin{equation}
  t = \kappa b [ 1- \exp ( -\gamma W/b)][1-\exp ( - \beta \Capi^{2/3})],
  \label{Film}
\end{equation}
with $\kappa \approx 0.119$, $\gamma \approx 0.038$ and $\beta \approx 8.58$.

\begin{figure} 
  \begin{center}
  \includegraphics[height=5.8cm]{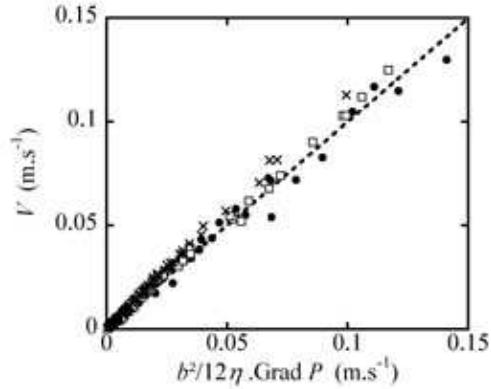}
  \caption{Velocity as a function of the applied pressure gradient for all viscosities (V02, V05, V10, V20 and V100) and different cell geometries: $\bullet$, $b$=1.43\mm \ $W$=4\cm; $\times$, $b$=0.75\mm \ $W$=4\cm; $\Box$, $b$=0.75\mm \ $W$=8\cm; $+$, $b$=0.25\mm \ $W$=4\cm.}
  \label{GDarcy}
  \end{center}
  
\end{figure}

In this way, we can thus test the validity of Darcy's law. Figure \ref{GDarcy} shows the velocity $V$ represented as a function of $(b/12\eta) \nabla P$  for the different cell geometries and viscosities used. The dashed line represents the linear relation with slope unity expected from (\ref{DarcyMoy}). We therefore conclude that the data are in excellent agreement with the classical Darcy's law. This result holds even for high velocities where a significant effect of the inertial forces is observed on the width of the fingers, as will be discussed below. We have thus shown that for the range of \Rey \ tested in this paper there is, as was anticipated above, no effect of inertia on Darcy's law when considering the uniform flow far away from the finger. Note that this does of course not mean that there are no corrections to the local Darcy's law near the finger tip. 

\subsection{Finger width} 

\subsubsection{Relative finger width as a function of the classical control parameter}

\begin{figure} 
  \begin{center}
  \includegraphics[height=5.8cm]{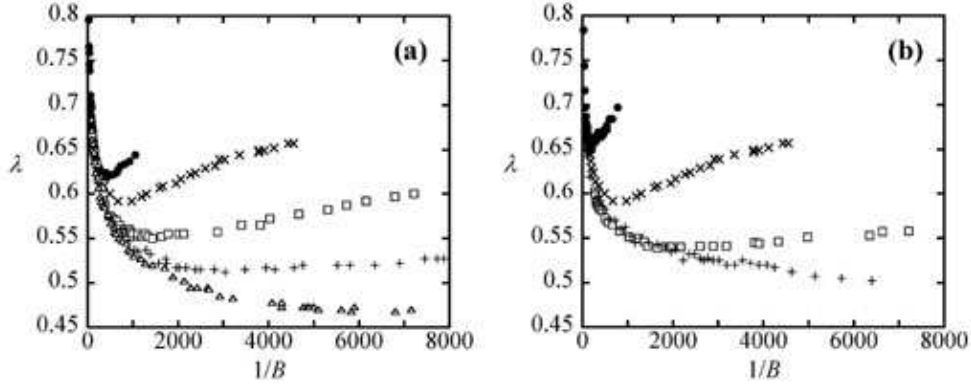}
  \caption{Results for the finger width $\lambda$ as a function of the classical control parameter \onB : (a) for the geometry with $b$=0.75\mm \ and $W$=4\cm \ and different fluids: $\bullet$, V02; $\times$, V05; $\Box$, V10; $+$ V20; $\triangle$, V100. (b) for the V05 fluid and different cell geometries: $\bullet$, $b$=1.43\mm \ $W$=4\cm; $\times$,~$b$=0.75\mm \ $W$=4\cm; $\Box$, $b$=0.75\mm \ $W$=8\cm; $+$, $b$=0.25\mm \ $W$=4\cm.  }
  \label{GonB}
  \end{center}
\end{figure}

Figures \ref{GonB}(a) and \ref{GonB}(b) represent the relative finger width as a function of the classical control parameter \onB \ when varying the viscosity of the fluid for a given geometry -$b$=0.75\mm \ $W$=4\cm, figure \ref{GonB}(a)- and when changing the geometry of the cell for a given fluid -silicon oil 47V05, figure \ref{GonB}(b). These figures show that, for low \onB \ one observes the classical decrease of the finger width with increasing \onB. However at a given value of \onB \ which is different for different configurations, an increase of the relative finger width is observed. This surprising observation systematically appears at high Reynolds number, and we conclude that it must be related to inertial effects. Indeed, for a given geometry, only the fluid of highest viscosity gives results that agree with the classical Saffman-Taylor instability. In addition, deviations from the classical results arise at smaller \onB \ for lower fluid viscosity. Finally, the data for a fixed viscosity but varying geometry (figure \ref{GonB}b) show that the increase of the finger width occurs for lower \onB \ for a thicker channel. All these observations agree with the suggestion that the increase in finger width with increasing velocity is due to inertial effects.
 
Comparing the data for a fixed gap thickness ($b$=0.75\mm) and two different channel widths ($W$=4 and $W$=8\cm) on figure \ref{GonB}(b), it follows that the crossover value of \onB \ also depends on the channel width $W$: it is observed to be smaller for smaller channel width. This is still consistent with an increase of the Reynolds numbers (\Rey \ and \Reym): at a given \onB \ and fixed $\eta$ and $b$, a decrease of the channel width $W$ leads to an increase of \Rey. This follows from the observation that $W^2U$ is fixed and consequently \Rey \ varies as $1/W$  (\Reym \ as $1/W^3$).

We conclude that due to inertia our experimental results deviate from the classical curves: we observe a regime of increasing finger width. This increase occurs at lower \onB \ for lower viscosity, larger gap thickness or smaller gap width. It is also important to note that strong inertial effects are observed already at velocities below 0.08\ms \ for all of our experiments and that even if strong changes in the behaviour are observed for the finger width, no deviations from the classical Darcy law is observed in this regime (figure~\ref{GDarcy}).

We will now study the deviation from the classical result as a function of the modified Reynolds number \Reym. For concreteness, in the following we will focus on the data obtained when varying the viscosity for a given geometry ($b$=0.75\mm, $W$=4\cm). The results however are general and apply to the data from other experiments as well. 

\subsubsection{Relative finger width as function of the modified Reynolds number \Reym}

\begin{figure} 
  \begin{center}
  \includegraphics[height=5.8cm]{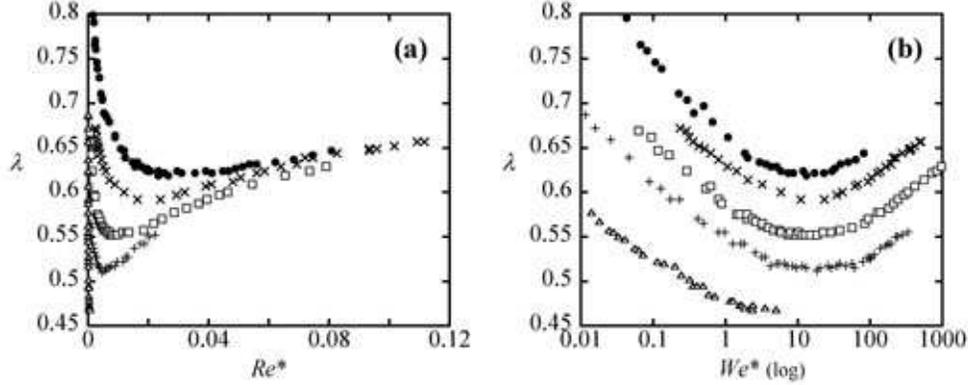}
  \caption{Results for the finger width $\lambda$ as a function of the modified Reynolds number \Reym \ (a) and of the modified Weber number \Webm \ (b), for the geometry with $b$=0.75\mm \ and $W$=4\cm \ and different fluids: $\bullet$, V02; $\times$, V05; $\Box$, V10; $+$ V20; $\triangle$, V100.}
  \label{GReWe}
  \end{center}
\end{figure}

In figure \ref{GReWe}(a), we plot the relative finger width as a function of the modified Reynolds number \Reym. The first observation is that the minimum of the curves that signals the deviation from the classical results, is not given by \Reym. On the other hand for high \Reym \ all the curves tend towards a single master curve: the behaviour of the finger width seems to be governed by \Reym \ only. Data in other configurations (not shown here) confirm the existence of a universal  $\lambda$-\Reym curve.

\subsubsection{Relative finger width as a function of the modified Weber number \Webm}

So far we can distinguish between two limiting cases. For low velocities, the results for the relative finger width fall on the universal curve of the classical Saffman-Taylor instability: they rescale with \onB. For high values velocity, a second universal curve exists and the data rescale with \Reym. It follows that the crossover between the two regimes is given by the modified Weber number, combining \Reym \ and \onB:
\begin{equation}
  \Webm=\Reym.\onB=\frac{\rho U^2W}{\gamma}=\frac{W}{b}\Web.
  \label{WeReonB}
\end{equation}

The experimental data support this conclusion. Figure \ref{GReWe}(b) depicts the relative finger width as a function of the modified Weber number \Webm. All experimental curves have a minimum located at the same value of \Webm \ at around \Webmc ~$\approx$~15 separating the two limiting behaviours.

Note that, although \Webm \ governs the crossover, we observe no regime where the finger width is given by a competition between capillary forces and inertia. In fact, when considering the dependence of the different forces on the velocity one finds that capillary forces scale as $U^0$, viscous forces as $U^1$ and inertial forces as $U^2$. Consequently, the dominating forces should at low velocity be capillary and viscous forces (control parameter \onB) and at high velocity, viscous forces versus inertia (control parameter \Reym). This simple argument therefore explains that as a function of the velocity there is no regime where the finger width is given by \Webm.

\subsubsection{Extension to a new global master curve}

It follows that the parameter \Webm \ ($=\Reym.\onB$), that can be seen as the ratio between \onB \ and 1/\Reym, gives the relative importance of the two parameters with a cross over given by the critical value \Webm ~$\approx$~15.

We can thus attempt to define a modified control parameter taking this crossover into account: 
\begin{equation}
  \onBm=\onB\left(\frac{1}{1+\Webm/\Webmc}\right).
  \label{DefonBm}
\end{equation} 

It is easily seen that this parameter tends to \onB \ for low \Webm \ (\Webm$<$\Webmc) and towards $\Webmc/\Reym$ for large \Webm \ (\Webm$>$\Webmc).

\begin{figure} 
  \begin{center}
  \includegraphics[height=5.8cm]{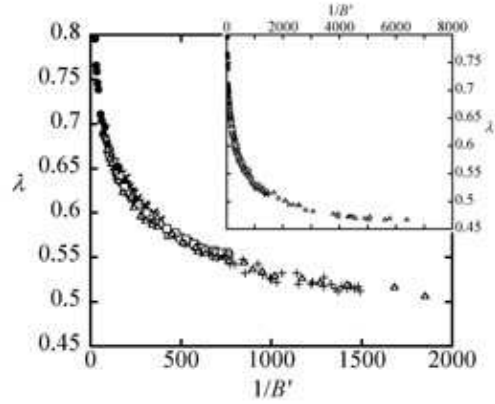}
  \caption{Results for the finger width $\lambda$ (same data as figure \ref{GonB}a) as a function of the modified control parameter \onBm, for the geometry with $b$=0.75\mm \ and $W$=4\cm \ and different fluids: $\bullet$,~V02; $\times$, V05; $\Box$, V10; $+$ V20; $\triangle$, V100. [Insert: large scale]}
  \label{GonBm}
  \end{center}
\end{figure}

Figure \ref{GonBm} shows the experimental data already shown on figure \ref{GonB}(a), however now $\lambda$ is plotted as a function of \onBm. Surprisingly the experimental data scale on a single universal curve when represented as a function of the modified control parameter. Moreover, and perhaps even more surprisingly this curve is identical to the result of the classical Saffman-Taylor instability.

\begin{figure} 
  \begin{center}
  \includegraphics[height=5.8cm]{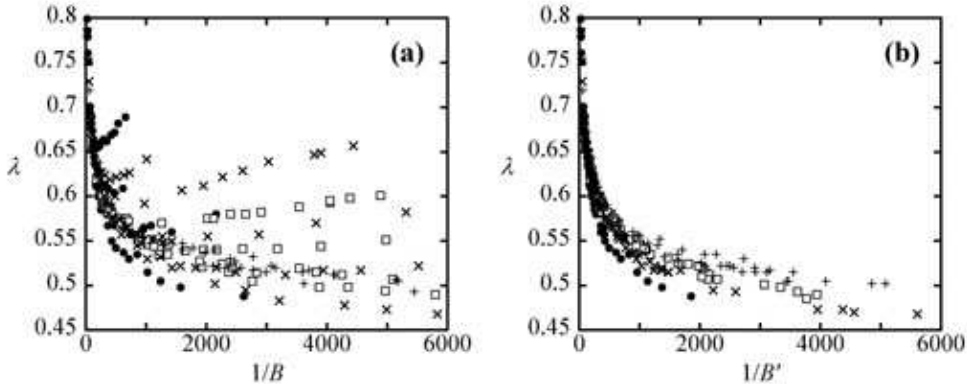}
  \caption{Results for the finger width $\lambda$ as a function of the classical control parameter \onB \ (a) and of the modified control parameter \onBm \ (b) for all viscosities (V02, V05, V10, V20 and V100) and different cell geometries: $\bullet$, $b$=1.43\mm \ $W$=4\cm; $\times$, $b$=0.75\mm \ $W$=4\cm; $\Box$, $b$=0.75\mm \ $W$=8\cm; $+$, $b$=0.25\mm \ $W$=4\cm.  }
  \label{GonBonBm}
  \end{center}
\end{figure}

The figures \ref{GonBonBm}(a) and \ref{GonBonBm}(b) summarize our results. On figure \ref{GonBonBm}(a), all geometries and fluids are depicted representing the relative finger width as a function of the classical control parameter \onB. One should emphasize that this data are obtained by varying not only the fluid viscosity but also the cell geometry by changing both, the channel width and thickness. On figure~\ref{GonBonBm}(b), the same data are plotted as a function of \onBm, our modified control parameter. We observe that the entire data set collapses very reasonably onto the single master curve when represented as a function of \onBm. Once again this curve is identical to the classical result of \cite{Saff58}.

So far we did not discuss the influence of the aspect ratio on the relative finger width. Even if the influence of the latter is small it might explain the fact we do observe slight deviations from the four different cell geometries (see figure \ref{GonBonBm}b). However when considering one single channel geometry the data do collapse (see figure \ref{GonBm}). If one also notes that these differences can already be observed where inertia is negligible, the conclusion must be that the slight residual differences are due to film effects.

The physical interpretation of our results is then the following. The modified control parameter \onBm \ gives the cross-over between \onB \ and \Reym. For small \Webm, \onB \ is the control parameter, and the main forces are surface tension and viscous forces, leading to a narrowing of the fingers width as viscous forces become more important for higher speeds. For higher velocities (\Webm$>$\Webmc), the main acting forces become viscous forces and inertia. The competition between these forces results in a widening of the finger width with increasing velocity. The observation is therefore that inertia tends to widen the fingers; this seems logically intuitively, as the inertia will tend to slow down the finger at a given flow rate, leading consequently to wider fingers. As the effect of the inertial forces is similar to that of the capillary forces in, the sense that both tend to widen the finger, and the classical Saffman-Taylor finger selection appears to have remained intact, one may attempt to  include the inertial forces in an effective surface tension. Indeed the modified control parameter \onBm \ can be written as the classical control parameter \onB \ by including an effective surface that is of the form:
\begin{equation}
  \gamma_{eff}=\gamma ( 1+\Webm/\Webmc).
  \label{gammod}
\end{equation}

\section{Some theoretical elements}

\begin{figure} 
  \begin{center}
  \includegraphics[height=3.5cm]{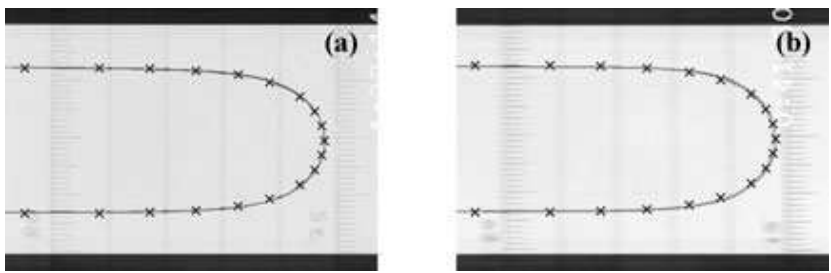}
  \caption{Snapshot of a finger: (a) without inertial effects (\Webm=5 $<\Webmc$, silicon oil V02, velocity $U$=5.2\cms, $b$=0.75\mm, $W$=4\cm); (b) with inertial effects (\Webm=30 $>\Webmc$, silicon oil V02, velocity $U$=12.5\cms, $b$=0.75\mm, $W$=4\cm). $\times$: Finger shape predicted without inertial effects by the theory of \cite{Pitt80}.}
  \label{shape}
  \end{center}
\end{figure}

In addition to the above, the experimental observations indicate that even if the finger width increases when increasing the velocity sufficiently the finger shape does not really change (see figure \ref{shape}).  All these observations suggest the possibility to introduce the inertial effects in a perturbative manner into the framework of the classical Saffman-Taylor treatment of the viscous fingering instability.

\subsection{Perturbation of the Darcy's law}

Modifications of Darcy's law have already been introduced in \S \ \ref{secInertia}. We will now consider the Euler-Darcy equation and thus a Darcy equation corrected by inertia, in the frame of the moving finger.

One starts from (\ref{Darcy2}) for the two-dimensional velocity field $\boldsymbol{u}(x,y,t)$ in the laboratory frame.
In the frame of the moving finger the problem is by definition stationary. Considering $\boldsymbol{u}(x,y,t)=\boldsymbol{u'}(x-Ut,y)+U\boldsymbol{e_x}$, we obtain for the velocity $\boldsymbol{u'}(x,y)$ in frame of reference of the moving finger:
\begin{equation}
  \rho \left( -\alpha U \frac{\partial\boldsymbol{u'}}{\partial x}+\beta U \frac{\partial\boldsymbol{u'}}{\partial x}+\beta\boldsymbol{u'}\bnabla \boldsymbol{u'}\right)=-\bnabla p-\frac{12\eta}{b^2}\boldsymbol{u'}.
  \label{Darcy2fing}
\end{equation}

Using the same scaling as before and omitting the~$^\ast$ (dimensionless symbols)  and the~$'$ (finger frame) for the variables, we find:
\begin{equation}
  \Reym \left( (\beta-\alpha) \frac{\partial\boldsymbol{u}}{\partial x}+\beta\boldsymbol{u}\bnabla \boldsymbol{u}\right)=-\bnabla p-\boldsymbol{u}.
  \label{Darcy2fing2}
\end{equation}

We assume that the flow remains a potential flow, i.e. $\boldsymbol{u}=\bnabla\phi$. This restriction to potential flow without vorticity is possible as long as the boundary conditions that apply to the finger and the walls are not modified. If this is the case, one can write:
\begin{equation}
  \phi=-\left[p+\Reym[(\beta-\alpha)u_x+\mbox{$\frac{1}{2}$}\beta u^2]+x+cst\right] \quad \mbox{and\ }\quad \Delta \phi =0.
  \label{DefPhi}
\end{equation}

Considering that along the finger and away from its tip, there are no inertial effects (the fluid is at rest in the laboratory frame, and $\lim_{x\to-\infty}\boldsymbol{u}=-\boldsymbol{e_x}$ in the finger frame) one should choose the constant equal to: $\Reym(\beta/2-\alpha)$.

The mechanical equilibrium of the interface requires the balance of the normal stress from both sides, given by  (\ref{DeltP}): $p=-\tilde{\gamma}/R$ ($R>0$), where $\tilde{\gamma}=(b/W)^2\gamma/(12\eta  U)=\onB^{-1}$  is the dimensionless surface tension and $R$ is the dimensionless radius of curvature.  Using this, one obtains the following boundary condition for $\phi$ at the interface:
\begin{equation}
  \phi_\Gamma=-\left[-\tilde{\gamma}/R+\Reym[(\beta-\alpha)u_x+\mbox{$\frac{1}{2}$}\beta u^2+\beta/2-\alpha]+x\right].
  \label{PhiInt}
\end{equation}

As the normal velocity at the interface $u_n$ is zero in the frame of the moving finger, the only remaining velocity component is the tangential one $u_t$ and we can use the notation of \cite{McLe81}: $\boldsymbol{u}=u_t\boldsymbol{e_t}=-q(\cos\theta\boldsymbol{e_x}+\sin\theta\boldsymbol{e_y})$ where $q$ varies from $0$ (at the tip of the finger) to $1$ (at its side) when $\theta$ varies from $-\upi/2$ to $0$. We can thus replace $u^2$ by $q^2$ and $-u_x$ by $q\cos\theta$.

As $q$ is mainly given by $\cos\theta$, one can write:
\begin{equation}
\phi_\Gamma=-x+\tilde{\gamma}\left[1/R+\Webm(\alpha-\beta/2)\sin^2\theta\right].
  \label{PhiInt2}
\end{equation}

Note that the last term is the Bernoulli correction. In all previous mentioned references \cite[see][]{Gond97, Ruye01, Plou02} $(\alpha-\beta/2)$ is positive, so this correction has the same sign as the curvature. It also vanishes at the sides of the finger. This shows that the effect of the inertial term is very similar to that of the capillary forces: the inertial forces should tend to increase the finger width, as was indeed observed experimentally.

Finally, rewriting (\ref{PhiInt2}), it follows that:
\begin{equation}
  \phi_\Gamma=-x+\hat{\gamma}(\theta)/R, \quad \mbox{with\ }\quad \hat{\gamma}(\theta)= \tilde{\gamma}\left[1+\Webm(\alpha-\beta/2)R(\theta)\sin^2\theta\right],
  \label{GamEff2}
\end{equation}
a form identical to (\ref{gammod}) obtained above by considering the modified control parameter \onBm \ with an effective surface tension.

\subsection{Numerical simulations and comparison to experimental data}

Of course the selection of the relative width of the finger can only be found by a sophisticated singular perturbation analysis. However, the relative finger width can be obtained numerically by a modification of the \cite{McLe81} method. We choose this method for a comparison with the experimental results and introduce the correction of \cite{Ruye01} in the numerics using (\ref{PhiInt}). We also introduce the effect of the wetting film by modifying in the value of the surface tension as done in (\ref{Film}).

\begin{figure} 
  \begin{center}
  \includegraphics[height=5.8cm]{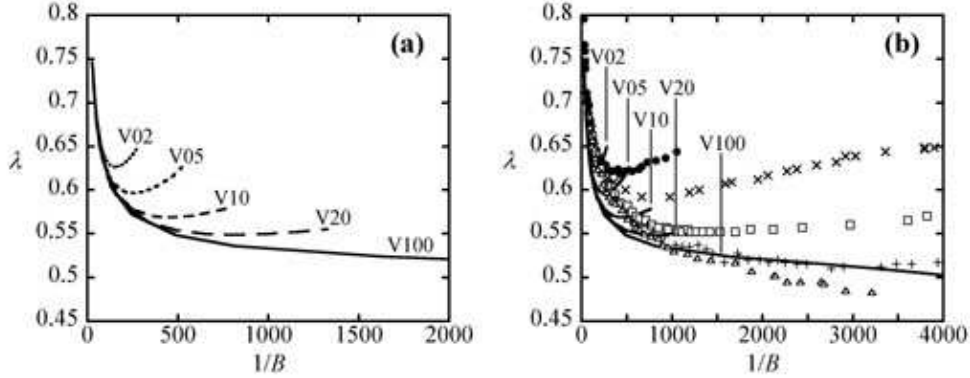}
  \caption{Finger width $\lambda$ as a function of the classical control parameter \onB \ for the geometry with $b$=0.75\mm \ and $W$=4\cm \ and different fluids: numerical simulations (a) and comparison between simulations and experimental results, experimental data: $\bullet$, V02; $\times$, V05; $\Box$, V10; $+$~V20; $\triangle$, V100; Numerical simulations: lines.}
  \label{Numeric}
  \end{center}
\end{figure}

If this is done, the numerical simulations confirm the simple argument pointed out above and show an increase of the relative finger width compared to the classical results (see figure \ref{Numeric}a) in agreement with the experiments. The observations from the numerical results are:
\begin{itemize}
\item
The inertial effects are stronger (i.e. appear for a smaller critical \onB) for less viscous fluids as well as for larger cell thickness or for smaller cell width.

\item
The minima of the relative width as a function of \Webm \ are around a unique value of \Webmc, however the numerical value of \Webmc \ differs between the experiments ($\approx$15) and the simulations ($\approx$2).
\end{itemize}

When comparing the simulations and the experiments (see figure \ref{Numeric}b) it turns out that they are in qualitative agreement but that the results are not identical.

When inertial effects are present we can characterize these effects by estimating a critical value of the control parameter \onBc \ and a critical relative width $\lambda_c$ at the minimum. It turns out that the numerics provide a rather good estimate for $\lambda_c$ but fall to give a correct value for \onBc \ which is found to be smaller in the numerical simulations than in the experiments. Another significant difference is that the increase of the $\lambda$-\onB \ curve is observed to be stronger in the simulations.

However, even for the case where inertia can be neglected (for most viscous fluid), there is a small but significant discrepancy between the numerical simulations and the experimental data. We believe this is due to the fact that our way of correcting for the wetting film is too simple. In consequence, it is clear that we can not expect perfect agreement between experimental data and simulations when adding corrections due to inertia. To quantitatively account for the experimental results, one has certainly to go back to the three-dimensional effects of the experiment which are expected to be important close to the finger tip over a length scale of order $b$. This is due to the existence of a three-dimensional structure of the flow which cannot be ignored in the vicinity of the finger: a film exists between the plate and the finger. \cite{Park84} and \cite{Rein85} have shown that it is nevertheless possible to reduce the problem to two dimensions when modifying the boundary conditions on the finger \cite[see also][]{BenA02}. However, they have also shown that reduction to two-dimensional is only possible if the parameter $\Web=\rho U^2 b/\gamma$ is small. For our problem, this is not the case and therefore a complete set of the corrected boundary conditions must be deduced from the full three-dimensional theory of Park, Homsy and Reinelt incorporating inertial effects in order to resolve the problem, which is largely beyond the scope of this paper.

\section {Summary and conclusion}

We have investigated the effect of inertia on the Saffman-Taylor instability. Inertial effects are found to become important for high Reynolds numbers and thus for fluids of low viscosity and for large plate spacing of the Hele-Shaw cell. For these situations one observes, upon increasing the velocity, first a classical regime with a decrease of the relative finger width and then a second and new regime in which the finger width increases. This second regime is due to the importance of inertia.
 
We introduced a modified Weber number \Webm \ which allowed us to explain the crossover between the two regimes. The transition is thus given by a critical modified Weber number \Webmc.
Below \Webmc, the classical regime of decreasing finger width is of course governed by the classical control parameter \onB, which is a modified capillary number. The finger width in this regime is thus given by the balance between capillary forces, which tend to widen the finger, and viscous forces, which tend to narrow the finger. With increasing velocity the viscous forces dominate over the capillary forces and one observes a narrowing of the finger.
For the second regime, above \Webmc, one observes on the contrary an increase of the finger width with increasing velocity. In this case the finger width is governed by a modified Reynolds number \Reym \ and thus by the balance between viscous forces and inertia. It turnsout that inertial forces tend to widen the finger. With increasing velocity inertia dominates the viscous forces and one consequently observes a widening of the fingers.
 
We have also shown that we can define a new control parameter \onBm, which takes the corrections due to inertia into account. This parameter tends towards \onB \ for low \Webm \ and is proportional to 1/\Reym \ for large \Webm. When plotting our data as a function of this empirical parameter they collapse onto a single universal curve which corresponds to the results for the finger width obtained for the classical Saffman-Taylor instability.

By only taking into account a modification of Darcy's law, some simple arguments and numerical simulations confirm all of these observations. However, the agreement between numerics and experiments is only qualitative. We believe this is due to the fact that the problem is certainly three-dimensional and one must consider the full three-dimensional theory of Park, Homsy and Reinelt incorporating inertial effects.

\begin{acknowledgments}
We thank Eric Cl\'ement, Mike Shelley and Laurent Limat for usefull discussions and Jos\'e Lanuza for valuable help with the experimental set-up.
\end{acknowledgments}


\end{document}